\documentclass[prb,preprint,showpacs,preprintnumbers,amsmath,amssymb]{revtex4}

\usepackage{graphicx}

\newcommand{\be}{\begin{equation}}
\newcommand{\ee}{\end{equation}}
\newcommand{\eq}[1]{Eq.~(\ref{#1})}
\newcommand{\fig}[1]{Fig.~\ref{#1}}
\def\beq{\begin{eqnarray}}
\def\eeq{\end{eqnarray}}
\def\beqa{\begin{eqnarray}}
\def\eeqa{\end{eqnarray}}

\def\bra{\langle}
\def\ket{\rangle}
\def\vq{{\bf q}}

\def\vk{{\bf k}}
\def\vQ{{\bf Q}}

\begin{document}

\title{Possible charge instabilities in two-dimensional doped Mott insulators}

\author{Mat\'{\i}as Bejas$^{*}$, Andr\'es Greco$^\dag$, and Hiroyuki Yamase$^\ddag$}
\affiliation{
{$^*$}The Abdus Salam International Centre for Theoretical Physics,
Strada Costiera 11, 34151 Trieste, Italy\\
{$^\dag$}Facultad de Ciencias Exactas, Ingenier\'{\i}a y Agrimensura and
Instituto de F\'{\i}sica Rosario (UNR-CONICET),
Av. Pellegrini 250, 2000 Rosario, Argentina\\		
{$^\ddag$}National Institute for Materials Science, Tsukuba 305-0047, Japan\\ 
{$^\ddag$}Max-Planck-Institute for Solid State Research, D-70569 Stuttgart, Germany
}

\date{\today}

\begin{abstract}
Motivated by the growing evidence of the importance of 
charge fluctuations in the pseudogap phase in 
high-temperature cuprate superconductors,  
we apply a large-$N$ expansion formulated in a path integral representation 
of the two-dimensional $t$-$J$ model on a square lattice. 
We study all possible charge instabilities of the paramagnetic state 
in leading order of the $1/N$ expansion. 
While the $d$-wave charge density wave (flux phase) 
becomes the leading instability for various choices of model parameters, 
we find that a $d$-wave Pomeranchuk (electronic nematic phase) instability 
occurs as a next leading one. 
In particular, the nematic state 
has a strong tendency to become inhomogeneous. 
In the presence of a large second nearest-neighbor 
hopping integral, the flux phase is suppressed and 
the electronic nematic instability becomes leading in a high doping region. 
Besides these two major instabilities, 
bond-order phases occur as weaker instabilities close to half-filling. 
Phase separation is also detected in a finite temperature 
region near half-filling. 
\end{abstract}


\maketitle

\section{Introduction}

The pseudogap (PG) phase in cuprate superconductors provides one of
the most active subjects on high-$T_c$ superconductivity.
The PG phase is characterized by highly anomalous properties\cite{timusk99,norman05} 
which are rather universal for all cuprate superconductors. 
One of the puzzling observations 
comes from angle-resolved photoemission spectroscopy (ARPES) 
measurements,\cite{damascelli03} which revealed arc-shaped disconnected 
Fermi surfaces,\cite{norman98}  called Fermi arcs, instead of a large Fermi surface.
In underdoped cuprates the PG opens below a temperature $T^*$, 
which is far above the superconducting transition temperature $T_{sc}$. 
Furthermore, in contrast to the behavior of $T_{sc}$, 
$T^*$ increases with decreasing doping in the underdoped region. 
The PG is very anisotropic along the Fermi surface. 
It has a maximal gap in the $(0,0)$-$(0,\pi)$ direction (antinodal direction)
and vanishes upon approaching the Brillouin zone diagonal (nodal direction), 
similar to the  $d$-wave superconducting gap. 

In spite of the consensus on the existence of the PG, its origin and
nature remain elusive. 
There are two major scenarios. 
One is  that the PG originates from
preformed pairs above $T_{sc}$.\cite{emery95,norman07}  
The other is that the PG is distinct from the
superconducting gap and associated with a certain order which competes 
with superconductivity, but 
both coexist at low temperature, leading to ``two gaps'' in 
the electronic spectrum.\cite{vishik10,kondo11,yoshida12} 
Several phenomenological models which are in favor of the two-gap scenario 
were already studied in various contexts, but invoking different orders,  
such as $d$-wave charge density wave ($d$CDW),\cite{chakravarty01} 
$d$-wave Fermi surface deformations,\cite{yamase09}  
charge density wave\cite{castellani95,becca96,hashimoto10,he11} 
including stripes,\cite{kivelson03,vojta09} 
phase separation (PS),\cite{emery93,castellani95,becca96} 
and others such as 
resonating-valence-bond-type charge order\cite{yang06}  
and loop-current order.\cite{varma0699} 

The $d$CDW is a flux phase, where 
orbital currents flow around each plaquette in a staggered pattern. 
The electronic spectrum in the flux phase has a gap with $d$-wave 
symmetry, the same as the superconducting gap symmetry. 
The flux phase was obtained in the large-$N$ approach to the $t$-$J$ model 
in various formalisms\cite{affleck88a,affleck89,morse91,cappelluti99,foussats04} 
and the presence of flux correlations was confirmed by the exact diagonalization.\cite{leung00}  
On the other hand, in the Hubbard model, 
the dynamical cluster approximation failed to detect 
static long-range order of the $d$CDW\cite{macridin04}  
whereas the variational cluster approximation 
showed that the $d$CDW is a metastable solution.\cite{lu11}  
Fluctuations associated with the $d$CDW 
can provide a route to address the PG. 
A perturbative analysis of the electron self-energy due to $d$CDW fluctuations 
catches many important features observed 
by ARPES, not only a PG and its associated Fermi arcs\cite{greco09,bejas11} but also 
a semiquantitative aspect of renormalization of 
the electron band dispersion in the PG phase.\cite{greco11} 

The $d$-wave Fermi surface deformations are driven by a 
$d$-wave Pomeranchuk\cite{pomeranchuk59} 
instability ($d$PI), leading to an electronic nematic state.\cite{kivelson98} 
In this state, an orientational symmetry of the systems is broken 
without breaking however translational invariance. The $d$PI was found 
in the slave-boson mean-field,\cite{yamase00} exact diagonalization,\cite{miyanaga06} 
and variational Monte Carlo\cite{edegger06} studies in the $t$-$J$ model, 
and also in the Hubbard model.\cite{metzner00,wegner02,buenemann12} 
The $d$PI itself does not become the leading instability in most of theoretical studies. 
However, it was pointed out that the models retain appreciable correlations of 
the $d$PI,\cite{yamase00,okamoto10} 
which then may lead to a giant response to a small $xy$ anisotropy. 
Such a giant response was actually observed in the PG region in 
YBa$_2$Cu$_3$O$_y$, which has a small anisotropy originating from 
the orthorhombic crystal structure, 
by neutron scattering\cite{hinkov07,hinkov08} and transport measurements.\cite{daou10} 
Theoretical studies for the former\cite{yamase06,yamase09} 
and the latter\cite{hackl09} confirmed that idea.

Charge-stripe order is extensively discussed 
for cuprates.\cite{kivelson03,vojta09} 
Since the charge-stripe order breaks both orientational and translational 
symmetry of the system, the stripe phase is also called an electronic 
smectic phase\cite{kivelson98} and has lower symmetry than the nematic phase.  
The experimental observation of charge order  
in La-based cuprates\cite{tranquada95} provides grounds to consider 
the stripe order. 
A charge-stripe solution was indeed obtained in the 
density matrix renormalization group (DMRG) study\cite{white98,white00} 
in the $t$-$J$ model. However in the presence of 
the second nearest-neighbor hopping integral the charge-stripe order turned out 
to be unstable in the $t$-$J$ model.\cite{white99,tohyama99}

PS is also another possible instability in the $t$-$J$ model.\cite{marder90,emery90}  
It is however still highly debated whether the model indeed shows the instability 
toward PS\cite{hellberg00,ivanov04} or not\cite{yokoyama96,white98,white00,hellberg99,shih98,putikka00,lugas06} 
in a parameter region realistic to cuprates. 
Although PS is in general suppressed by long-range Coulomb forces, 
strong charge fluctuations in the proximity to PS can be important 
and responsible for anomalous properties in the PG 
phase and superconductivity.\cite{emery93} 
In fact, the proximity to PS plays an important role 
to generate a singular interaction between electrons at zero momentum transfer 
as shown in the infinite-$U$ Hubbard Holstein model.\cite{castellani95,becca96} 
When long-range Coulomb interactions are added, 
the singularity shifts to 
a finite momentum transfer, leading to an incommensurate charge density wave  
similar to stripes.\cite{castellani95,becca96}

Theoretically it is believed that the 
two-dimensional (2D) $t$-$J$ and Hubbard models  
contain the main ingredients for describing cuprates,\cite{anderson87}  
i.e., antiferromagnetism at
zero doping, a metallic state at finite doping, and a strong tendency to
$d$-wave superconductivity. 
Given that various charge instabilities are invoked to address 
the PG, and also other anomalous properties in cuprates, 
it is interesting to study 
what kind of charge instabilities are favored in the 2D $t$-$J$ model 
by treating all possibilities on equal footing in a controllable scheme. 

In this paper, we analyze the 2D $t$-$J$ model in terms of Hubbard
operators by including the nearest-neighbor Coulomb interaction $V$ 
to avoid a subtle feature of PS; our main results are not 
affected by the presence of $V$. 
We apply a large-$N$ expansion formulated in a path integral 
representation.\cite{foussats00,foussats02,foussats04}  
In this approach the two spin components are
extended to $N$ and an expansion in powers of
the small parameter $1/N$ is performed, 
providing a controllable scheme 
without a perturbative expansion in any model parameter.  
In addition, different kind of instabilities can be studied on equal footing, 
allowing us to perform a stability analysis on
all possible charge instabilities already at leading order. 
We find that the $t$-$J$ model shows tendencies to the flux 
and electronic nematic state in a wide doping region. 
In particular, the nematic state has a strong tendency to become inhomogeneous. 
Close to half-filling, bond-order phase (BOP) and PS are also obtained. 

In the next section, we first provide a brief summary of our theoretical scheme 
and then explain the most important charge instabilities, 
$d$CDW, $d$PI, BOP, and PS. Our results are presented in Sec.~III 
and are discussed by comparing with literature in Sec.~IV. 
Implications for cuprates are also discussed in the same section. 

\section{Theoretical framework}
\subsection{Large-\boldmath{$N$} approach to the \boldmath{$t$-$J$-$V$} model}
In a previous paper,\cite{foussats04}  
a large-$N$ expansion for the $t$-$J$-$V$ model was formulated in terms of 
a path integral representation for the Hubbard $X$-operators. 
For the sake of a self-contained presentation, 
we first summarize the formalism. 

The $t$-$J$-$V$ model is described by the following Hamiltonian, 
\be\label{Hc}
H = -\sum_{i, j,\sigma} t_{i\,j}\tilde{c}^\dag_{i\sigma}
\tilde{c}_{j\sigma}
+ J \sum_{\bra i,\, j\ket} (\vec{S}_i \cdot \vec{S}_j-\frac{1}{4} n_i n_j)
+ V \sum_{\bra i,\, j\ket} n_i n_j \, ,
\ee
where $t_{i\,j} = t$ $(t')$ is the hopping integral between the 
first (second) nearest-neighbor sites on a square lattice;  
$J$ and $V$ are the exchange interaction
and the Coulomb repulsion, respectively, between the nearest-neighbor sites. 
The main role of the $V$-term in the present study is to suppress the tendency toward 
PS 
while, in other works,\cite{zeyher98,gazza99} the $V$-term was included to investigate 
its effect on superconductivity. 
$\tilde{c}^\dag_{i\sigma}$ and $\tilde{c}_{i\sigma}$ are 
the creation and annihilation operators of electrons 
with spin $\sigma$ ($\sigma=\downarrow$,$\uparrow$),  respectively, 
under the constraint that 
the double occupancy of electrons is excluded 
at any site $i$. $n_i$ is the electron density operator. 

The electron and spin operators are connected to Hubbard operators\cite{hubbard63}  
via  $\tilde{c}^\dag_{i \sigma}=X_i^{\sigma 0}$, 
$\tilde{c}_{i \sigma}=X_i^{0 \sigma}$, 
$S_i^+=X_i^{\uparrow \downarrow}$, 
$S_i^-=X_i^{\downarrow \uparrow}$, 
and $n_i=X_i^{\uparrow \uparrow}+X_i^{\downarrow \downarrow}$. 
The operators $X_i^{\sigma 0}$ and $X_i^{0 \sigma}$ are called fermionlike, 
whereas the operators $X_i^{\sigma \sigma'}$ and
$X_i^{00}$ are called bosonlike; $X_i^{00}$ will be introduced later [\eq{eq:v1}]. 
After writing Hamiltonian (1) in terms of the Hubbard operators,  
we extend the spin degree of freedom to $N$ channels and obtain 
the Hamiltonian in the large-$N$ formalism, 
\be
H = - \frac{1}{N}\sum_{i, j, p}\; t_{i j} X_{i}^{p 0}X_{j}^{0p} 
+ \frac{J}{2N} \sum_{\bra i,j\ket, pp'} (X_{i}^{p p'}X_{j}^{p' p} - X_{i}^{p p} X_{j}^{p' p'}) 
+ \frac{V}{N}\sum_{\bra i,j\ket, p p'} X_{i}^{p p} X_{j}^{p' p'}
-\mu\sum_{i,p}\;X_{i}^{p p} \label{eq:H} \,.
\ee
The spin index $\sigma$ is extended to a new index $p$, which 
runs from $1$ to $N$. In order to obtain a finite theory in the $N$-infinite limit,  
$t$, $t'$, $J$ and $V$ are rescaled as
$t/N$, $t'/N$, $J/N$ and $V/N$, respectively. 
The chemical potential $\mu$ is introduced in \eq{eq:H}. 

In the path integral formulation our Euclidean Lagrangian reads 
\begin{eqnarray}
L_E =  \frac{1}{2}
\sum_{i, p}\frac{({\dot{X_{i}}}^{0 p}\;X_{i}^{p 0}
+ {\dot{X_{i}}}^{p 0}\;
X_{i}^{0 p})} {X_{i}^{0 0}} + H
\label{lagrangian}
\end{eqnarray}
with the following two additional constraints, 
\begin{eqnarray}
X_{i}^{0 0} + \sum_{p} X_{i}^{p p} - \frac{N}{2}=0 \; , \label{eq:v1}
\end{eqnarray}
and
\begin{eqnarray}
X_{i}^{p p'} - \frac{X_{i}^{p 0} X_{i}^{0p'}}{X_{i}^{0 0}}=
0 \;, \label{eq:v2}
\end{eqnarray}
which are imposed on the path integral via two $\delta$-functions.
In \eq{lagrangian}, ${\dot{X_{i}}}^{p 0}=\partial_\tau {X_{i}}^{p 0}$ 
and $\tau$ is the euclidean time, namely $\tau={\rm i}t$. 
Equation~(\ref{eq:v1}) is the $N$-extended completeness condition.  
The form of the kinetic term in the Lagrangian $L_E$, as well as 
the constraint \eq{eq:v2}, comes from the requirement that  
the $X$-operators should fulfill their commutation rules. 
For details we refer to 
Ref.~\onlinecite{foussats00}. 
In the path integral approach we associate Grassmann and usual
bosonic variables with fermionlike and bosonlike $X$-operators,
respectively. 

We now discuss the main steps needed to introduce a large-$N$
expansion.\cite{foussats02,foussats04} 
First the $V$-term in the Hamiltonian 
is written in terms of $X_{i}^{0 0}$ by using \eq{eq:v1}. 
We then eliminate the bosonic variables $X^{p p'}$ 
by implementing the $\delta$-function associated with \eq{eq:v2}. 
The completeness condition [\eq{eq:v1}] is imposed by 
introducing Lagrange multipliers $\lambda_i$. 
We write $X_{i}^{0 0}$ and $\lambda_i$  in terms of
static mean-field values, $r_0$ and $\lambda_0$, and fluctuation fields, 
$\delta R_i$ and $\delta \lambda_i$, 
\begin{eqnarray}
X_{i}^{0 0} &=& N r_{0}(1 + \delta R_{i}) \nonumber \\
\lambda_{i} &=&\lambda_{0}+ \delta{\lambda_{i}} \,.
\label{lambdai}
\end{eqnarray}
In addition, we introduce the following fermion fields\cite{misc-fandX} defined by 
\begin{eqnarray}\label{fdag}
f^{\dag}_{i p} &=& \frac{1}{\sqrt{N r_{0}}}X_{i}^{p 0} \nonumber \,,\\
f_{i p} &=& \frac{1}{\sqrt{N r_{0}}}\;X_{i}^{0 p}\,.
\end{eqnarray}
The exchange interaction is then described by four fermion fields, which 
are decoupled through a Hubbard-Stratonovich transformation 
by introducing a field associated with a bond variable, 
\be
\Delta_{ij} = J \sum_{p} \frac{f^{\dag}_{j p} f_{i p}}{ \sqrt{(1 + \delta R_{i})
(1 + \delta R_{j})}} \,. 
\ee
The field $\Delta_{ij}$ is parameterized by 
\be
\Delta_i^{\eta}=\Delta(1+r_i^\eta+iA_i^\eta)\,,
\label{staticDelta}
\ee  
where $r_i^{\eta}$ and $A_i^{\eta}$ correspond to the real and imaginary parts of the 
fluctuations of the bond variable,
respectively, and $\Delta$ is a static mean-field value. 
The index $\eta$ takes two values associated with the bond directions
 ${\eta}_{1}=(1,0)$ and ${\eta}_{2}=(0,1)$ on a square lattice. 
After expanding $1/(1+\delta R)$ in powers of $\delta R$, 
we obtain an effective Lagrangian, which can be written in terms of 
a six-component boson field 
\be
\delta X^{a} = (\delta
R\;,\;\delta{\lambda},\; r^{{\eta}_{1}},\;r^{{\eta}_{2}}
,\; A^{{\eta}_{1}},\;
A^{{\eta}_{2}})\,,
\ee 
the fermions $f_{p}$, and their interactions.

From the quadratic part for fermions we obtain an electronic 
propagator in the paramagnetic phase, 
\begin{eqnarray}\label{G0}
G({\bf k}, {\rm i}\nu_{n}) = \frac{1}{{\rm i}\nu_{n} -\varepsilon_{\vk}}\,.
\end{eqnarray}
Here ${\vk}$ and ${\rm i}\nu_{n}$ are the momentum and fermionic  Matsubara frequency, respectively, and the electronic dispersion $\varepsilon_{\vk}$ is 
\begin{eqnarray}\label{Ek}
\varepsilon_{\vk} = -2(t r_0+\Delta) (\cos k_{x}+\cos k_{y})-
4t' r_0 \cos k_{x} \cos k_{y} - \mu \,. 
\end{eqnarray}
Here $\lambda_0$ in \eq{lambdai} was absorbed in the chemical potential $\mu$.

From the completeness condition [\eq{eq:v1}] $r_0$ is equal to $\delta/2$,  
where $\delta$ is the hole doping rate away from half-filling. 
The field $\Delta$ is given  
by the expression 
\beq{\label {Delta}}
\Delta = \frac{J}{4N_s} \sum_{\vk, \eta} \cos(k_\eta) n_F(\varepsilon_\vk) \; , 
\eeq
where $n_F$ is the Fermi function and $N_s$ is the total 
number of lattice sites. For a given doping, 
$\mu$ and $\Delta$ are determined self-consistently by solving \eq{Delta} and 
\be
(1-\delta)=\frac{2}{N_s} \sum_{\vk} n_F(\varepsilon_\vk)\,.
\ee

The quadratic part for $\delta X^{a}$ 
defines a $6\times6$ bare bosonic propagator 
$D^{(0)}_{ab}(\vq,\mathrm{i}\omega_n)$, which 
after Fourier transformation reads, 
\begin{widetext}
\begin{equation} \label{D0inverse}
[D^{(0)}_{ab}(\vq,\mathrm{i}\omega_n)]^{-1} = N 
\left(
\begin{array}{llllll}
\frac{\delta^2}{2}\left(V-\frac{J}{2}\right)
[\cos(q_x) + \cos(q_y)] & \frac{\delta}{2} & 0 & 0 & 0 & 0\\
\frac{\delta}{2} & 0 & 0 & 0 & 0 & 0\\
0 & 0 & \frac{4\Delta^2}{J} & 0 & 0 & 0\\
0 & 0 & 0 & \frac{4\Delta^2}{J} & 0 & 0\\
0 & 0 & 0 & 0 & \frac{4\Delta^2}{J} & 0\\
0 & 0 & 0 & 0 & 0 & \frac{4\Delta^2}{J}
\end{array}
\right) \; ,
\end{equation}
\end{widetext}
where ${\vq}$ and ${\rm i}\omega_{n}$ are the momentum and bosonic 
Matsubara frequency, respectively. 
The quantity $D^{(0)}_{ab}(\vq,\mathrm{i}\omega_n)$ 
describes all possible types of bare charge susceptibilities. 
The bare susceptibilities are renormalized already at leading order 
to become dressed ones,  which are given by the Dyson equation 
\be
D^{-1}_{ab}(\vq,\mathrm{i}\omega_n)
= [D^{(0)}_{ab}(\vq,\mathrm{i}\omega_n)]^{-1} - \Pi_{ab}(\vq,\mathrm{i}\omega_n)\,.
\label{dyson}
\ee
Following the diagrammatic rules in Ref.~\onlinecite{foussats04}, 
the $6\times6$ boson self-energies $\Pi_{ab}$ are computed as 
\begin{widetext}
\begin{eqnarray}\label{eq:Piab}
&& \Pi_{ab}(\vq,\mathrm{i}\omega_n)
            = -\frac{N}{N_s}\sum_{\vk} h_a(\vk,\vq,\varepsilon_\vk-\varepsilon_{\vk-\vq}) 
            \frac{n_F(\varepsilon_{\vk-\vq})-n_F(\varepsilon_\vk)}
                                  {\mathrm{i}\omega_n-\varepsilon_\vk+\varepsilon_{\vk-\vq}} 
            h_b(\vk,\vq,\varepsilon_\vk-\varepsilon_{\vk-\vq}) \nonumber \\
&& \hspace{25mm} - \delta_{a\,1} \delta_{b\,1} \frac{N}{N_s}
                                       \sum_\vk \frac{\varepsilon_\vk-\varepsilon_{\vk-\vq}}{2}n_F(\varepsilon_\vk) \; .
\end{eqnarray}
\end{widetext}
The prefactor $N$ in front of the right hand side of Eq.~(17) comes from 
the sum over the $N$ channels of $p$. 
Thus, the $6\times6$ boson self-energies $\Pi_{ab}$ are 
of the same order as $[D^{(0)}_{ab}]^{-1}$ [see Eq.~(15)]. 
In Eq.~(17) $h_a$ is an effective six-component interaction vertex which comes 
from the interaction terms between bosonic and fermionic fields 
derived from the effective Lagrangian. 
The explicit expression for $h_a$  is given by 
\begin{widetext}
\begin{align}
 h_a(\vk,\vq,\nu) =& \left\{
                   \frac{2\varepsilon_{\vk-\vq}+\nu+2\mu}{2}+
                   2\Delta \left[ \cos\left(k_x-\frac{q_x}{2}\right)\cos\left(\frac{q_x}{2}\right) +
                                  \cos\left(k_y-\frac{q_y}{2}\right)\cos\left(\frac{q_y}{2}\right) \right];1;
                 \right. \nonumber \\
               & \left. -2\Delta \cos\left(k_x-\frac{q_x}{2}\right); -2\Delta \cos\left(k_y-\frac{q_y}{2}\right);
                         2\Delta \sin\left(k_x-\frac{q_x}{2}\right);  2\Delta \sin\left(k_y-\frac{q_y}{2}\right)
                 \right\} \; . 
\end{align}
\end{widetext}

From the $N$-extended 
completeness condition [\eq{eq:v1}] we see that the charge operator 
$X^{00}$ is $O(N)$, while the operators $X^{pp}$ are $O(1)$. 
Consequently, the $1/N$ approach emphasizes   
the effective charge interactions. 
In fact, while in leading order  
charge susceptibilities  contain collective effects, they 
enter the spin susceptibilities in the next-to-leading order. 
Similarly, superconductivity appears 
in the next-to-leading order.\cite{cappelluti99,zeyher98}   
Therefore, instabilities of the paramagnetic phase 
are expected only, in leading order, in the charge sector. 

In leading order, our formalism agrees with the $1/N$ slave-boson 
formalism.\cite{morse91}  
However, in the present approach the fermion variables $f_{ip}$ are proportional to the
$X$-operators [\eq{fdag}] and should not be confused with
the spinons in the slave-boson approach. In addition, $\delta R$ [\eq{lambdai}] 
is proportional to charge fluctuations and not related to holons. 
Since the $X$-operators are treated as fundamental objects, 
problems associated with fluctuations of gauge fields 
in the slave-boson approach\cite{lee06} are avoided. 
Our formalism was also checked to yield results consistent with 
the exact diagonalization\cite{merino03,bejas06} 
as well as results in another formalism of 
the $1/N$ expansion in leading order.\cite{cappelluti99}

\subsection{Instabilities of the paramagnetic phase} 
An instability of the paramagnetic phase is signaled by 
the divergence of the static susceptibilities defined 
by $D_{ab}(\vq,\mathrm{i}\omega_n)$ for a continuous phase transition. 
Therefore we study eigenvalues and eigenvectors 
of the matrix $[D_{ab}(\vq, \rm{i}\omega_n)]^{-1}$ at $\rm{i}\omega_n=0$. 
When an eigenvalue crosses zero at a given doping rate, temperature $T$, 
and momentum $\vq$, an instability occurs toward a phase characterized 
by the corresponding eigenvector.
We have found five instabilities associated with eigenvectors $V^a$ 
explained below.   

\begin{figure}[htp]
\centering
\includegraphics[width=8cm]{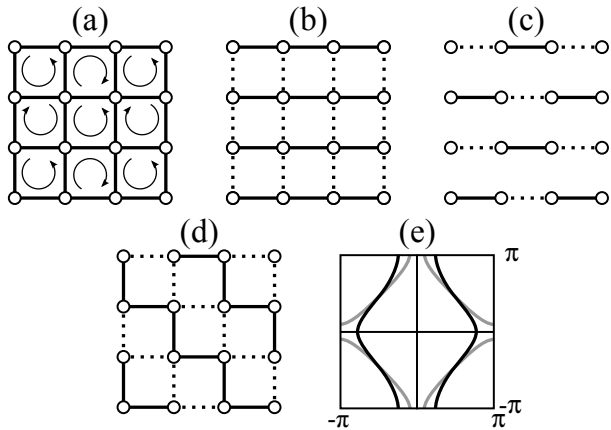}
\caption{Sketch of various phases appearing in our work. 
Commensurate orders for (a) $d$CDW, (b) $d$PI, 
(c) BOP$_x$, and (d) BOP$_{xy}$ in real space. 
The commensurate $d$PI has a momentum $\vq=(0,0)$ 
whereas the commensurate $d$CDW, BOP$_x$, and BOP$_{xy}$ 
have $\vq=(\pi,\pi)$. Solid and dashed lines in (b)-(d) 
represent the strong and weak bonds, respectively.
(e) Fermi surface deformations (black line) 
associated with the commensurate $d$PI; the original Fermi surface 
is sketched by gray lines. 
}
\label{fig1}
\end{figure}

a) $V^a =\frac{1}{\sqrt{2}} (0,0,0,0,1,-1)$, which corresponds to the freeze of  
the imaginary parts of the bond variable [\eq{staticDelta}]. 
The pure imaginary contribution to the hopping term generates 
a net magnetic flux in each plaquette, leading to 
the instability toward the flux or $d$CDW phase as already 
found previously.\cite{affleck88a,affleck89,morse91,cappelluti99,foussats04}
The commensurate flux phase is characterized by the 
modulation vector $\vq=(\pi,\,\pi)$ 
and describes staggered circulating currents as sketched in \fig{fig1} (a),    
whereas the incommensurate phase is characterized by $\vq\ne(\pi,\,\pi)$. 

b) $V^a = \frac{1}{\sqrt{2}}(0,0,1,-1,0,0)$, which corresponds to the freeze of 
the real  parts of the bond variable. 
This eigenvector corresponds to the commensurate [$\vq=(0\,,0)$] or 
incommensurate  [$\vq\ne(0\,,0)$] instability toward the $d$PI phase. 
The commensurate phase is sketched in \fig{fig1} (b) in real space. 
In momentum space it leads to Fermi surface deformations where 
the Fermi surface expands along the $k_y$ direction and shrinks 
along the $k_x$ direction [\fig{fig1} (e)], 
or vice versa if the bond along the $y$ direction would become stronger than the $x$ direction. 
While the commensurate $d$PI has been discussed since 2000,\cite{yamase00,metzner00} 
an incommensurate $d$PI starts to be discussed 
very recently.\cite{metlitski10,metlitski10a,holder12,husemann12} 

c) $V^a = (0,0,1,0,0,0)$ ($(0,0,0,1,0,0)$), which correspond to the freeze of 
the third (fourth) component and describe the instability toward 
the BOP$_x$ (BOP$_y$).\cite{cappelluti99,foussats04,morse91} 
The corresponding modulation vector turns out to be $\vq=(\pi,\pi)$  or 
very close to it. The commensurate BOP$_x$, namely with $\vq=(\pi,\pi)$, 
is sketched in \fig{fig1} (c) 
whereas the BOP$_y$ with $\vq=(\pi,\pi)$ 
is obtained by rotating \fig{fig1} (c) by 90$^{\circ}$.  

d) $V^a = \frac{1}{\sqrt{2}} (0,0,1,1,0,0)$, which corresponds to the freeze of 
both third and fourth components simultaneously. 
The modulation vector is $\vq=(\pi,\pi)$ or 
very close to it, as in the case of the BOP$_{x}$ and BOP$_{y}$. 
We refer to this instability as the BOP$_{xy}$. 
The commensurate BOP$_{xy}$ with $\vq=(\pi,\pi)$ 
is sketched in \fig{fig1} (d). For simplicity, we also use the phrase BOP 
when we do not have to distinguish between BOP$_x$, BOP$_y$, and BOP$_{xy}$.

e) $V^a = (1,0,0,0,0,0)$, which corresponds to the freeze of 
charge fluctuations $\delta R$ and describes 
the instability toward PS for $\vq=(0,0)$ and a charge-density-wave phase, 
including stripes, for a finite $\vq$. 
A finite $\vq$ instability, however, was not detected in the 
present study.

In general, eigenvectors of $[D_{ab}(\vq, \rm{i}\omega_n)]^{-1}$ can have a non-zero 
value in each component. 	
However we checked that the inner product
between an eigenvector of $[D_{ab}(\vq, \rm{i}\omega_n)]^{-1}$ and $V^a$ 
becomes larger than $0.99$ at the corresponding critical point. 
 
\subsection{Effective susceptibilities}
While numerical results presented in this paper are 
computed from the full susceptibility \eq{dyson}, 
it is instructive to extract an effective susceptibility associated with 
each instability explained in the previous section 
by discarding the interactions with other modes contained in $D_{ab}(\vq, \rm{i}\omega_n)$. 

The usual charge-charge correlation function is written as 
\begin{eqnarray}\label{chi}
\chi^c_{ij}(\tau)=\frac{1}{N} \sum_{p,q} \bra T_\tau X_i^{pp}(\tau)
X_j^{qq}(0)\ket \; .
\end{eqnarray}
Using the completeness condition [\eq{eq:v1}] and the relation between
$X^{00}_i$ and $\delta R_i$ [\eq{lambdai}], $\chi^c_{ij}$ can be written  
in Fourier space,\cite{foussats04}
\begin{eqnarray}\label{chi-c}
\chi^c({\bf q},{\rm i}\omega_n)=
-N {\left( \frac{\delta}{2} \right)}^2 D_{11}({\bf q},{\rm i}\omega_n)\,.
\end{eqnarray}
Thus the charge-charge correlation function is just the component $(1,1)$ of the $D_{ab}$.
Note that the factor $N$ in front of the right hand side of \eq{chi-c} 
shows that 
charge fluctuations are of $O(1)$ since $D_{ab} \propto 1/N$ as seen in \eq{D0inverse}. 

The susceptibility of the $d$CDW is obtained by 
focusing on the sector $a, b=5, 6$ of the matrix $D_{ab}^{-1}$. 
We obtain 
\begin{eqnarray}
\chi_{d{\rm CDW}}(\vq,{\rm i}\omega_n)=
 [(8/J) \Delta^2-\Pi_{d{\rm CDW}}(\vq,{\rm i}\omega_n)]^{-1}\,,
\label{chi-flux}
\end{eqnarray}
where $\Pi_{d{\rm CDW}}(\vq,{\rm i}\omega_n)$ is the electronic polarizability of 
the $d$CDW and is given by 
\begin{eqnarray}
\Pi_{d{\rm CDW}}(\vq, {\rm i}\omega_n) = - \frac{1}{N_{s}}\;
\sum_{\vk}\; \gamma_{d{\rm CDW}}^2(\vq,\vk) \frac{n_{F}(\epsilon_{\vk + \vq}) 
- n_{F}(\epsilon_{\vk})} 
{\epsilon_{\vk + \vq} - \epsilon_{\vk}-{\rm i} \omega_n}\,,
\label{PiCDW}
\end{eqnarray}
with a form factor 
$\gamma_{d{\rm CDW}}(\vq,\vk)=2 \Delta [\sin(k_x+q_x/2)-\sin(k_y+q_y/2)]$.

Similarly, the susceptibility of the $d$PI is obtained from the sector 
$a, b=3, 4$ of the matrix $D_{ab}^{-1}$:   
\begin{eqnarray}
\chi_{d\rm PI}(\vq,{\rm i}\omega_n)= [(8/J) \Delta^2-\Pi_{d\rm PI}(\vq,{\rm i}\omega_n)]^{-1}
\label{chi-dPI}
\end{eqnarray}
and the electronic polarizability  of the $d$PI reads 
\begin{eqnarray}
\label{Pi-dPI}
\Pi_{d\rm PI}(\vq, {\rm i}\omega_n) = - \frac{1}{N_{s}}\;
\sum_{\vk}\; \gamma_{d\rm PI}^2(\vq,\vk) \frac{n_{F}(\epsilon_{\vk + \vq}) 
- n_{F}(\epsilon_{\vk})} 
{\epsilon_{\vk + \vq} - \epsilon_{\vk}-{\rm i}\omega_n} \,,
\end{eqnarray}
with a form factor 
$\gamma_{d\rm PI}(\vq,\vk)=2 \Delta [\cos(k_x+q_x/2)-\cos(k_y+q_y/2)]$. 

For the case of the BOP$_x$ we focus on the sector $a=b=3$ and obtain 
\begin{eqnarray}
\chi_{{\rm BOP}_x}(\vq,{\rm i}\omega_n)= 
[(4/J)\Delta^2 -\Pi_{{\rm BOP}_x}(\vq,{\rm i}\omega_n)]^{-1}\,,
\label{chi-BOP}
\end{eqnarray}
where the electronic polarizability is given by 
\begin{eqnarray}\label{Pi-BOP}
\Pi_{{\rm BOP}_x}(\vq, {\rm i}\omega_n) = - \frac{1}{N_{s}}\;
\sum_{\vk}\; 4 \Delta^2 \cos^2 (k_x+q_x/2) \frac{n_{F}(\epsilon_{\vk + \vq}) 
- n_{F}(\epsilon_{\vk})} 
{\epsilon_{\vk + \vq} - \epsilon_{\vk}-{\rm i}\omega_n}\,. 
\end{eqnarray}
For the case of $a=b=4$, i.e., BOP$_y$, 
the form factor in \eq{Pi-BOP} is replaced by $\cos^2( k_y+q_y/2)$.  
It is easily seen in \eq{Pi-BOP} that the BOP$_{x}$ and BOP$_{y}$ 
instabilities occur simultaneously, but with a different modulation vector: 
suppose $\vq=(q_x,q_y)$ for the BOP$_{x}$, then $\vq=(q_y,q_x)$ 
for the BOP$_{y}$. While we will not present results for the BOP$_{y}$, 
it should be understood that the instability of the  BOP$_{y}$ also exists. 

The susceptibility associated with BOP$_{xy}$ is given by the same equation 
as \eq{chi-dPI}, except that 
the form factor $\gamma_{d{\rm PI}}$  in \eq{Pi-dPI} is replaced by 
$\gamma_{{\rm BOP}_{xy}} (\vq,\vk)=2 \Delta [\cos(k_x+q_x/2)+\cos(k_y+q_y/2)]$. 

The form factor $\gamma_{d\rm CDW}(\vq,\vk)$  [$\gamma_{d\rm PI}(\vq,\vk)$] 
has a $\vk$ dependence of $\cos k_x-\cos k_y$ at $\vq=(\pi,\pi)$ [$\vq=(0,0)$], 
which indicates the $d$-wave character of the instability. 
Note that the $d$PI and $d$CDW belong to different 
eigenspace and are not connected with each other by changing the momentum $\vq$. 

While the terminology of the $d$PI itself makes sense when a modulation vector 
is close to $\vq=(0,0)$, we may consider formally a large $\vq$ 
in Eqs.~(\ref{chi-dPI}) and (\ref{Pi-dPI}). 
The $d$PI is then connected with the BOP$_{x}$ and  BOP$_{y}$ 
when $\vq$ is located along the direction of $(\pi,0)$-$(\pi,\pi)$ or 
$(0,\pi)$-$(\pi,\pi)$. Suppose $\vq'=(\pi,q_y)$, we can easily find 
\be
\Pi_{d\rm PI}(\vq', {\rm i}\omega_n) =
\Pi_{{\rm BOP}_x}(\vq', {\rm i}\omega_n) + 
\Pi_{{\rm BOP}_y}(\vq', {\rm i}\omega_n)\,,
\ee
by noting that 
\be
\frac{1}{N_{s}} \sum_{\vk}\; \sin k_x \cos (k_y + q_y /2) \frac{n_{F}(\epsilon_{\vk + \vq'}) 
- n_{F}(\epsilon_{\vk})} 
{\epsilon_{\vk + \vq'} - \epsilon_{\vk}-{\rm i}\omega_n}=0\,.
\ee 
We thus obtain 
\be
\chi_{d\rm PI}^{-1}(\vq',{\rm i}\omega_n)= 
\chi_{{\rm BOP}_x}^{-1}(\vq',{\rm i}\omega_n) + 
\chi_{{\rm BOP}_y}^{-1}(\vq',{\rm i}\omega_n) \,.
\label{dPI-BOP}
\ee
In particular, when $\vq'$ is equal to $\vQ\equiv (\pi,\pi)$, \eq{dPI-BOP} is reduced to 
\be
\chi_{d{\rm PI}}(\vQ, {\rm i}\omega_n)=\frac{1}{2}\chi_{{\rm BOP}_x}(\vQ, {\rm i}\omega_n)\,,
\label{dPI-BOP-pi}
\ee 
because 
$\chi_{{\rm BOP}_x}(\vQ, {\rm i}\omega_n)=\chi_{{\rm BOP}_y}(\vQ, {\rm i}\omega_n)$. 
Similarly, we can also obtain 
\be
\chi_{d{\rm PI}}(\vQ, {\rm i}\omega_n)=\chi_{{\rm BOP}_{xy}}(\vQ, {\rm i}\omega_n)\,.
\label{dPI-BOPxy}
\ee
Hence when the static BOP susceptibility diverges at $\vq=(\pi,\pi)$, 
the $d$PI susceptibility also diverges simultaneously at the same momentum 
unless it already diverges at a different momentum.  
In fact, the $d$PI with $\vq=(\pi,\pi)$ is equivalent to the 
BOP$_{xy}$ and is interpreted as superposition of the 
BOP$_x$ and BOP$_y$ as seen in \fig{fig1} (d).

\section{Results}
We choose the parameters, $J/t=0.3$ and $V/t=0.5$. We set $t=1$, and 
all quantities with dimension of energy are in units of $t$. 
Irrespective of the presence of the $V$-term, 
our theory catches intrinsic charge instabilities in the 2D $t$-$J$ model 
such as $d$CDW, $d$PI, and BOP, which are driven by the $J$-term.
We compute the full susceptibility \eq{dyson} for various choices of $t'$ 
by assuming the paramagnetic phase 
and determine possible charge instabilities in the plane of 
hole density $\delta$ and temperature $T$. 
At half-filling an analytical solution is  obtained and 
the $d$CDW, dPI, and BOP have the same onset temperature, 
$T_c=J/8=0.0375t$, at which the static field $\Delta$ [\eq{staticDelta}] also sets in. 
Away from half-filling ($\delta > 0.004$) our computation is fully numerical. 
Since we determine critical lines by studying the susceptibility, 
the transition is continuous. In other words, a possibility of a first order 
transition is not considered in the present analysis. 

\subsection{Results for \boldmath{$t'=0$}}
\begin{figure}[htp]
 \centering
 \includegraphics[width=8cm]{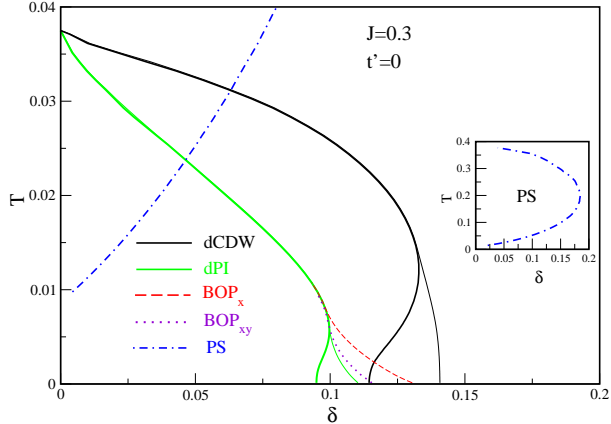}
 \caption{(Color online) Critical temperature versus doping rate 
for $d$CDW, $d$PI, BOP$_x$, BOP$_{xy}$, and PS for $t'=0$ and $J=0.3$. 
Thick (thin) lines describe commensurate (incommensurate) transitions. 
The critical line for PS is shown in a larger 
scale of $T$ in the inset.}
\label{fig2}
\end{figure}

Figure~\ref{fig2} shows the phase diagram for $t'=0$.
As mentioned in Sec.~II.B, five different types of charge instability are 
found: $d$CDW, $d$PI, BOP$_x$, 
BOP$_{xy}$, and PS. 
The instability toward the commensurate $d$CDW, namely with $\vq=(\pi,\pi)$, 
occurs in a wide doping region. The transition temperature decreases 
gradually with increasing hole density and exhibits reentrant behavior at low $T$ 
in the region $0.12 \lesssim \delta \lesssim 0.14$. 
However, near $\delta \approx 0.14$  an incommensurate [$\vq\ne(\pi,\pi)$] $d$CDW instability 
occurs below $T\approx 0.015$ and its critical doping rate is higher than that of 
the commensurate $d$CDW. Hence the resulting critical line of the $d$CDW 
follows the outer line, i.e., the thin line at low $T$ and the thick line for high $T$.

For $t'=0$ 
the commensurate $d$PI with $\vq=(0,0)$ has the same onset temperature as 
the commensurate BOP$_x$ and BOP$_{xy}$, namely with $\vq=(\pi,\pi)$; 
this reason will be explained in the last paragraph in the present section. 
The transition lines exhibit reentrant behavior at low $T$.   
However, an incommensurate $d$PI  
emerges at low $T$
and preempts the reentrant line, extending the region of $d$PI. 
BOP$_x$ and BOP$_{xy}$ also exhibit an incommensurate instability 
at low $T$ and preempt their commensurate instabilities. 
Furthermore the degeneracy between BOP$_x$ and BOP$_{xy}$ is lifted 
via an incommensurate transition. 
While the critical doping rate for BOP becomes  higher than for $d$PI at low $T$, 
this result occurs only for a small $t'$ and, as will be shown below, 
the opposite occurs in the presence of a realistic $t'$ for cuprates.  

The system also exhibits PS  at low doping. 
The inset of Fig. 2 shows the PS line in a larger temperature scale. 
We see that PS appears with decreasing temperature, 
but with further cooling down it goes back to the 
paramagnetic phase. As a result, PS occurs only 
in an intermediate temperature region. This peculiar reentrant behavior 
was also found in Ref.~\onlinecite{koch04} in the Hubbard model. 
The region of PS shrinks with increasing $V$ and also by introducing $t' (<0)$, 
as will be discussed in the subsection D.

The phase diagram in \fig{fig2} should not 
be interpreted in such a way that the $d$CDW is unstable against 
the $d$PI or BOP at low $T$ or low $\delta$, 
because we perform a stability analysis 
in the paramagnetic phase. 
Rather, \fig{fig2} indicates a hierarchy for different charge instabilities, i.e., 
the outer the critical line is, the stronger the tendency toward 
the corresponding instability is.

\begin{figure}[htp]
 \centering
 \includegraphics[width=12cm]{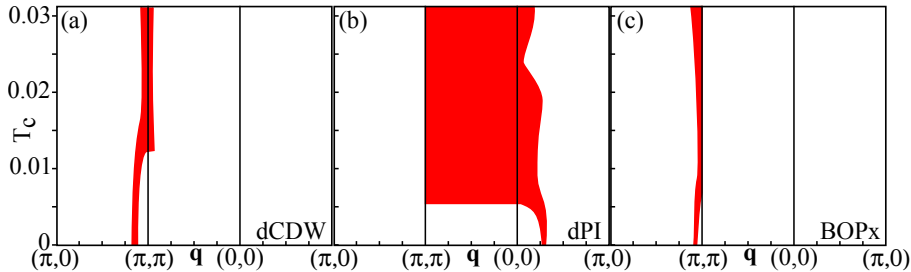}
\caption{(Color online) 
Modulation vectors for $d$CDW (left panel), $d$PI (middle panel), and 
BOP$_x$ (right panel) along the corresponding outer critical line 
in \fig{fig2}. For each critical temperature, the critical doping rate can be 
read off from \fig{fig2}.  Because of symmetry,  the results along the 
direction of $(0,\pi)$-$(\pi,\pi)$ are the same as those along  
the $(\pi,0)$-$(\pi,\pi)$ direction for the $d$CDW and $d$PI. 
}
\label{fig3}
\end{figure}

It requires highly accurate numerics to determine precisely a modulation vector $\vq$ 
of each order along its outer critical line because of a rather flat structure of the susceptibility 
in momentum space, especially for the $d$PI. 
Therefore considering our achieved numerical accuracy we present in \fig{fig3}   
modulation vectors of each instability, for which the absolute value of the 
corresponding eigenvalue of 
$D_{ab}^{-1}(\vq,0)$ becomes less than $10^{-4}t$ on its outer critical line. 
The width of such a $\vq$ region, at a fixed temperature, implies 
how sharp the susceptibility is in momentum space. 
Since modulation vectors of each instability are computed along its outer critical line, 
each critical temperature shown in \fig{fig3} corresponds to a certain critical doping rate, 
which can be read off from \fig{fig2}. 
Although \fig{fig3} is presented only along the axis $(\pi,0)$-$(\pi,\pi)$-$(0,0)$-$(\pi,0)$, 
we scanned the  whole $\vq$ region of the Brillouin zone and checked numerically 
that instabilities indeed occur along that axis.

\vspace{5mm}

In \fig{fig3} (a) we show the result for the $d$CDW along its outer critical line in \fig{fig2}. 
At high critical temperature (i.e., low critical doping rate) the instability occurs at $\vq = (\pi,\pi)$,
and with lowering temperature the modulation vector shifts 
from $(\pi,\pi)$ and becomes incommensurate. 
In \fig{fig3} (c) we plot the corresponding modulation vector of the BOP$_x$. 
At low $T$, the modulation vector $\vq$ slightly shifts 
from $(\pi,\pi)$ and the BOP$_x$ becomes incommensurate, as in the case of the $d$CDW.  
At high $T$, $\vq$ is located at $(\pi,\pi)$, but in contrast to the case of the 
$d$CDW, the $\vq$ region is not extended on the side of the direction of 
$(\pi,\pi)$-$(0,0)$. This is because the eigenvector of the BOP$_x$ does not exist there, 
instead, the eigenvector of the full susceptibility [\eq{dyson}] 
changes to that corresponding to the $d$PI. 
A modulation vector of the BOP$_{xy}$ appears only along the axis 
$(\pi,\pi)$-$(0,0)$. Its $T_c$ dependence is very similar to that of BOP$_x$, but 
the labels $(\pi,0)$ and $(0,0)$ in \fig{fig3} (c) 
should be replaced by $(0,0)$ and $(\pi,0)$, respectively.

The corresponding result for the $d$PI is shown in \fig{fig3} (b), which looks 
very different from those for the $d$CDW and BOP. 
In fact, for $t'=0$, the static electronic polarizability of the $d$PI 
has a special feature, which was already noted in Ref.~\onlinecite{morse91} 
in a different context. 
To see this we rewrite \eq{Pi-dPI} in a different form, 
\be
\Pi_{d\rm PI}(\vq, 0) = - \frac{1}{N_{s}}\;
\sum_{\vk}\;   n_{F}(\epsilon_{\vk})\left[
\frac{\gamma_{d\rm PI}^2(\vq,\vk)}{\epsilon_{\vk}-\epsilon_{\vk+\vq}} 
+\frac{\gamma_{d\rm PI}^2(-\vq,\vk)}{\epsilon_{\vk}-\epsilon_{\vk-\vq}} 
\right]\,.
\ee
When $\vq$ lies along the diagonal direction $\vq \parallel (q,q)$, 
we find after some algebra 
\be
\Pi_{d\rm PI}(\vq, 0) = \frac{8 \Delta^2}{(t\delta+2\Delta) N_{s}}\;
\sum_{\vk}\;   n_{F}(\epsilon_{\vk})\cos\frac{k_x+k_y}{2}
\tan\frac{k_x-k_y}{2}\sin\frac{k_x-k_y}{2}\,, 
\label{d-bubble}
\ee
that is, the static $d$PI susceptibility [\eq{chi-dPI}] 
does not depend on $\vq$ for any momentum along the diagonal direction. 
This result holds for any carrier density and any temperature. 
Therefore, if the $d$PI takes place for a vector $\vq$ in the diagonal direction, 
the susceptibility diverges simultaneously at all $\vq$ along the diagonal direction. 
The full susceptibility [\eq{dyson}] actually shows that feature in \fig{fig3} (b). 
Furthermore,  this flat feature of the susceptibility extends more away from the diagonal direction. 
The $\vq$ region along $(0,0)$-$(\pi,0)$ shrinks at $T_c \approx 0.024$ in \fig{fig3} (b), 
which results from the proximity to PS, as will be discussed in the subsection~C. 
While the susceptibility is always flat along the diagonal direction of $\vq$, 
the susceptibility shows a peak at a modulation vector along 
$(0,0)$-$(\pi,0)$ at low $T$. 
The $\vq$ region has a sharp boundary at $(\pi,\pi)$ in \fig{fig3} (b) and 
the $d$PI does not have any possible modulation vector 
along the $(\pi,0)$-$(\pi,\pi)$ direction. 
This is because 
the eigenvector corresponding to the $d$PI is not realized along 
$(\pi,0)$-$(\pi,\pi)$, instead,  the BOP$_x$ eigenvector appears there. 
This property may be understood also 
in terms of the effective susceptibilities. 
Equation~(\ref{dPI-BOP}) indicates that if $\chi_{d\rm PI}^{-1}(\vq',0)$ becomes 
zero, either $\chi_{{\rm BOP}_x}^{-1}(\vq',0)$ or $\chi_{{\rm BOP}_y}^{-1}(\vq',0)$ 
should be already negative,  
since in general $\chi_{{\rm BOP}_x}$ is not equal to $\chi_{{\rm BOP}_y}$ 
for a momentum along $(\pi,0)$-$(\pi,\pi)$, except for 
$\vq'=(\pi,\pi)$ where both $\chi_{{\rm BOP}_x}^{-1}(\vq',0)$ and 
$\chi_{{\rm BOP}_y}^{-1}(\vq',0)$ can become zero simultaneously.   
Therefore a possible instability of the $d$PI along $(\pi,0)$-$(\pi,\pi)$ 
is replaced by the BOP$_x$.

The $\vq$-independence of $\Pi_{d{\rm PI}}$ along the diagonal direction 
leads to another special feature. As we mentioned at the end of Sec.~II, 
the onset temperature of the BOP with $\vq=(\pi,\pi)$ is the same as that of the $d$PI 
with $\vq=(\pi,\pi)$ [Eqs.~(\ref{dPI-BOP-pi}) and (\ref{dPI-BOPxy})]. 
Therefore the onset temperature of 
the commensurate BOP becomes the same as that of the 
$d$PI with $\vq={\bf 0}$, as shown in \fig{fig2}.

\subsection{Results for finite \boldmath{$t'$}}
\begin{figure}[tp]
 \centering
 \includegraphics[width=8cm]{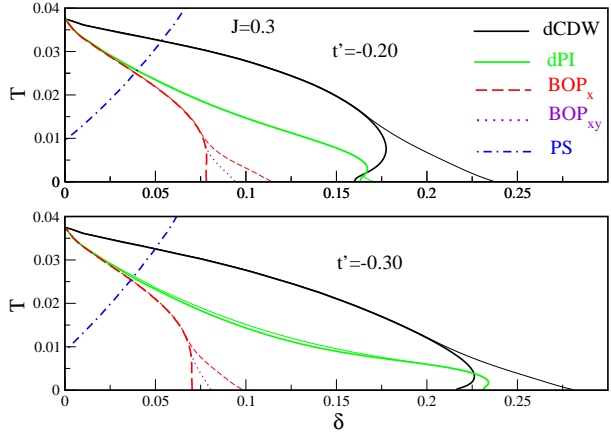}
 \caption{(Color online) Critical temperature and doping rate 
for $t'=-0.20$ (upper panel) and $t'=-0.30$ (lower panel). 
The notation is the same as \fig{fig2}. 
}
 \label{fig4}
\end{figure}

The degeneracy between the $d$PI and BOP seen in \fig{fig2} 
is lifted by introducing $t'$. 
The upper and lower panels in \fig{fig4} show 
the results for $t'=-0.20$ and $-0.30$, respectively. 
While the BOP instability is always restricted to a lower doping region,  
the $d$PI becomes favorable in a wider doping region. 
Near half-filling the $d$PI and BOP are still almost degenerate because,    
as seen from \eq{Ek}, the hopping integral $t'$ is renormalized to be 
$t' r_0 \propto t' \delta$ and becomes irrelevant close to half-filling. 
The BOP$_x$ and BOP$_{xy}$ are always degenerate as far as 
they exhibit a commensurate instability. 
Their degeneracy is lifted when their modulation 
vector becomes incommensurate at low temperature. 

As shown in \fig{fig4}, the doping region of the commensurate $d$CDW instability 
is extended by the presence of $t'$ and 
an incommensurate $d$CDW becomes dominant 
at high $\delta$ and low critical temperature. 
On the other hand, PS is suppressed by introducing $t'$. 
The critical line for PS bends back to zero doping for high $T$ (not shown) in a way similar 
to the case for $t'=0$ (inset of \fig{fig2}). 

\begin{figure}[tp]
 \centering
 \includegraphics[width=12cm]{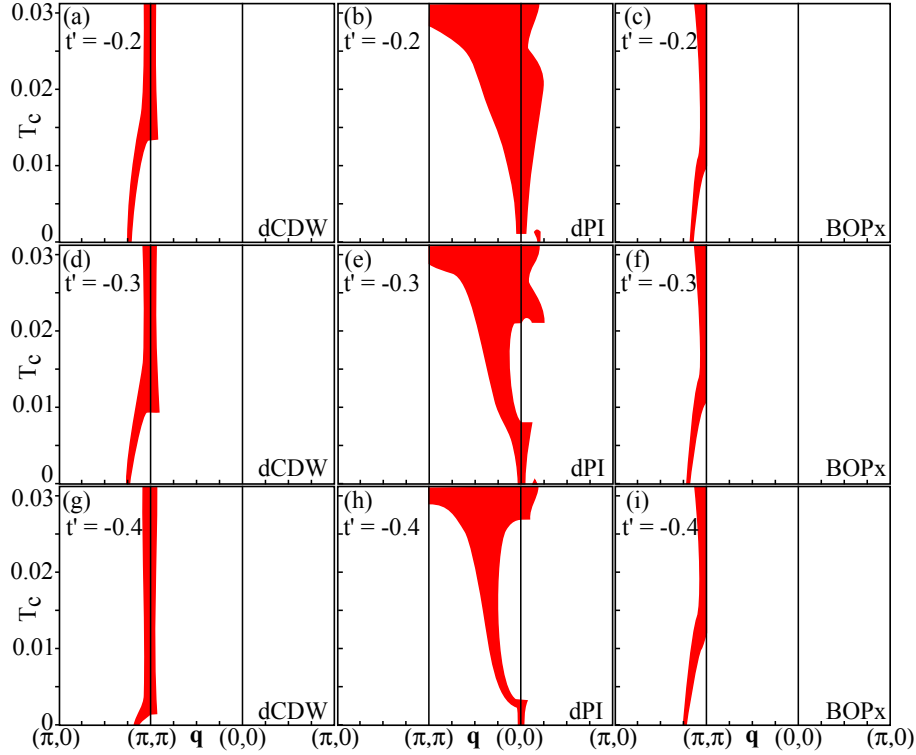}
\caption{(Color online) Modulation vectors for $d$CDW (left panels), $d$PI (middle panels), 
and BOP$_x$ (right panels) for $t'=-0.20$ (upper row), $t'=-0.30$ (middle row),
and $t'=-0.40$ (lower row) along the outer critical line of the corresponding order 
for each $t'$ in \fig{fig4} and \fig{fig6} (middle).  
}
 \label{fig5}
\end{figure}

The modulation vector of each instability is shown in the upper 
and middle row in \fig{fig5} for $t'=-0.20$ and $-0.30$, respectively, 
along the corresponding outer critical line in \fig{fig4}. 
Both $d$CDW [Figs.~\ref{fig5} (a) and (d)] and BOP$_x$ [Figs.~\ref{fig5} (c) and (f)] 
show an instability 
at $\vq=(\pi,\pi)$ for high critical temperature, and becomes incommensurate for low critical temperature. 
These features are the same as those seen in Figs.~\ref{fig3} (a) and (c). 
Results of BOP$_{xy}$ are essentially the same as those of BOP$_{x}$, 
but the labels $(\pi,0)$ and $(0,0)$ should be interchanged in Figs.~\ref{fig5} (c) and (f).

Figs.~\ref{fig5} (b) and (e) show results for the $d$PI. 
They are very different from \fig{fig3} (b), 
except for a region of high critical temperature near $T_c\approx 0.03$, 
namely the doping region $0< \delta_c \lesssim 0.02,$ where 
the effect of $t'$ becomes irrelevant. 
For $t'=-0.20$ [\fig{fig5} (b)], as the critical temperature decreases, 
the $\vq$ region shrinks around $\vq=(0,0)$ and 
the $d$PI tends to become commensurate. 
Close to zero temperature, however, 
a tendency toward an incommensurate $d$PI appears 
in the $(0,0)$-$(\pi,0)$ direction. 
This incommensurate feature is also seen more clearly for $t'=0$ at low $T$ [\fig{fig3} (b)], 
and disappears quickly with increasing $t'$. It becomes 
nearly invisible for $t'=-0.30$ in our temperature scale. 
For $t'=-0.30$,  
the $\vq$ region shrinks first around $\vq=(0,0)$ with 
decreasing the critical temperature. 
In the intermediate temperature range, $0.021 \gtrsim T_c \gtrsim 0.008$, 
the modulation vector becomes slightly incommensurate along the diagonal.
This deviation from the commensurate vector is also barely visible 
in the bottom panel of \fig{fig4},  where the incommensurate 
$d$PI line separates very slightly from the commensurate $d$PI in the corresponding 
temperature region. For $T_c \lesssim 0.008$, the $d$PI becomes fully commensurate.

\begin{figure}[tp]
 \centering
 \includegraphics[width=12cm]{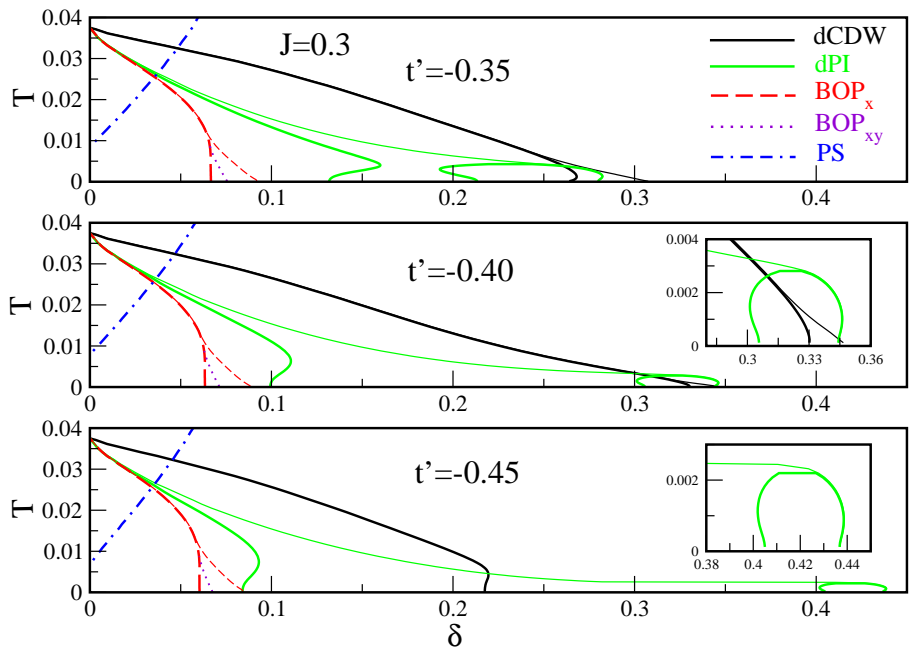}
 \caption{(Color online) The same plot as \fig{fig4}, but for different choices of $t'$: 
 $t'=-0.35$ (top panel), $t'=-0.40$ (middle panel), and $t'=-0.45$ (bottom panel). 
 In the latter two cases the phase diagram in a high doping region is 
 magnified in the inset.   
          }
 \label{fig6}
\end{figure}

The phase diagram close to half-filling 
does not depend essentially on a choice of $t'$.  
In fact, the critical lines for the 
BOP and PS do not change much even for a further larger $t'$. 
However, we find that tendencies toward $d$CDW and $d$PI have strong $t'$ 
dependence. 
In \fig{fig6} we present the phase diagram for $t'=-0.35$, $-0.40$, and $-0.45$. 
First we focus on the $d$CDW for $t'=-0.35$ and $-0.40$. 
The instability extends to a higher doping region with increasing $t'$ 
and the commensurate $d$CDW tends to become more favorable even at low $T$.  
The doping region of the $d$CDW, however,  
starts to decrease quickly for $|t'| > 0.41$, as can be seen in the result for $t'=-0.45$.

In contrast to the $d$CDW, the outer critical line for the $d$PI 
extends to higher doping with increasing $t'$. 
For $t'=-0.35$, an incommensurate $d$PI becomes dominant 
in a wide doping region ($0.05 \lesssim \delta \lesssim 0.25$). 
While the commensurate $d$PI is realized for 
$\delta \lesssim 0.16$ and $\delta \gtrsim  0.19$, 
it does not occurs between these two doping region. 
This feature is more evident for $t'=-0.40$ [\fig{fig6} (middle panel)]. 
In a wide doping region ($0.10 \lesssim \delta \lesssim 0.30$) 
the commensurate $d$PI does not occur 
and only an incommensurate $d$PI is possible. 
However in a high doping region ($0.30 \lesssim \delta \lesssim 0.35$), 
the commensurate $d$PI shows up again; 
see the inset in  \fig{fig6} (middle panel). 
These peculiar features of the $d$PI are due to the presence of the van Hove singularity around 
$\delta = 0.33$, where the $d$-wave weighted density of states,\cite{yamase05}  
which is defined by $\lim_{\vq \rightarrow {\bf 0}} \Pi_{d\rm PI}(\vq,0)$ in \eq{Pi-dPI},  
is enhanced, favoring the instability at $\vq=(0,0)$. 
In an intermediate doping region, 
the $d$-wave density of states 
is suppressed and the $d$PI with a finite $\vq$ 
becomes more favorable. 
Closer to half-filling, however, the $d$-wave density of states is again 
enhanced because of narrowing the band width upon approaching half-filling, 
leading to a recovery of the commensurate 
$d$PI for $\delta \lesssim 0.10$. 
For $t'=-0.45$ these features are more emphasized. 
Figure~\ref{fig6} (bottom panel) shows that 
the commensurate $d$PI occurs both, close to half-filling and 
around van Hove filling ($\delta=0.42$), and these two regions are connected 
by the critical line of an incommensurate $d$PI. 
The commensurate $d$PI close to half-filling and 
in a high doping region was also found in the slave-boson mean-field theory.\cite{yamase00} 

In \fig{fig5} we compare the modulation vector of each instability for 
$t'=-0.20$, $-0.30$, and $-0.40$. 
With increasing $t'$, the $d$CDW tends to be more commensurate 
even at low critical temperature. 
For the $d$PI, on the other hand, a large $t'$ tends to favor 
an incommensurate modulation along 
the diagonal direction of the Brillouin zone in an intermediate range of a critical 
temperature, and the commensurate $d$PI can be realized only at high and low critical temperature, 
corresponding to a doping region close to half-filling and around van Hove filling, 
respectively. The modulation vectors of the BOP$_x$ (and also BOP$_{xy}$) 
do not depend much on a choice of $t'$.

\subsection{Mutual interaction among different modes}
Because of the renormalization of the bosonic propagators due to the coupling 
to electronic bubbles [\eq{dyson}], one would expect in general some coupling 
among different modes. 
This effect actually appears for modulation vectors of the $d$PI. 
There is a dip on the side of the region of 
$(0,0)$-$(\pi,0)$ at $T_c\approx 0.024$ in \fig{fig3} (b), and 
$T_c \approx 0.025$ and $0.026$ in Figs.~\ref{fig5} (b) and (e), respectively. 
This dip occurs near the temperature where the critical lines of the $d$PI and PS 
cross each other (see Figs.~\ref{fig2} and \ref{fig4}). 
Moreover, we checked that the dip in question does not appear in results obtained 
from the effective susceptibility of the $d$PI [\eq{chi-dPI}]. 
Therefore a tendency toward PS plays a role in the suppression of the 
incommensurate feature of the $d$PI along $(0,0)$-$(\pi,0)$. 

Except for the above feature, we checked that our results (Figs.~\ref{fig2}-\ref{fig6})  
are nearly the same as those determined by the effective susceptibilities 
[Eqs.~(\ref{chi-c}), (\ref{chi-flux}), (\ref{chi-dPI}), and (\ref{chi-BOP})]. 
In this sense, the coupling among different bosonic fluctuations is rather weak 
at least in leading order.

\subsection{Effect of the \boldmath{$V$}-term and stability of phase separation} 
We checked that the results for $d$CDW, $d$PI, and BOP are almost intact 
for different choices of $V$ $(\geq 0)$ 
and  that an additional instability such as the usual checkerboard charge density wave 
does not occur at least for $V \leq 1$ for 
any doping rate.\cite{foussats04,misc-foussats04} 
Furthermore, the reentrant critical line of PS (inset of \fig{fig2}) is a robust feature. 
However, it is a subtle issue whether PS actually occurs at $T=0$. 
The result depends on choices of $V$, $t'$, and $J$. 
For $t'=0$ and $J=0.3$, we found that PS at $T=0$ occurs for $\delta \lesssim 0.08$ $(0.025)$ 
at $V=0$ $(0.1)$ and vanishes already for $V\gtrsim 0.2$. 
When $t'$ is introduced, PS is strongly suppressed even at $V=0$, 
for example, it occurs at $T=0$ only for $\delta \lesssim 0.01$  for $t'=-0.35$. 
A smaller $J$ also suppresses PS, and for the special case of $J=0$ 
no PS is observed at $T=0$ for any $V$ $(\geq 0)$ and $t'$ $(\leq 0)$.

\section {Conclusion} 
Applying a large-$N$ expansion formulated in the path integral representation 
of the $t$-$J$ model, we have analyzed all possible 
charge instabilities of the paramagnetic phase, and have elucidated  
the phase diagram in the doping and temperature plane for 
a sequence of $t'$.
We have found that $d$CDW, $d$PI, BOP, and PS are the most important 
charge instabilities. 
The first two instabilities appear in a wide doping region. 
The $d$CDW usually becomes the leading instability and the $d$PI 
occurs as a next leading one with a strong 
tendency to become incommensurate. 
In the presence of a large $t'$, however, 
we have found that the $d$PI becomes the leading instability 
in a high doping region.  
Considering the high complexity of the $t$-$J$ model, it 
is beyond the scope of the present study to address 
which charge instability would become ultimately the leading one. 
Rather, the present stability analysis on charge instabilities of the paramagnetic state 
was motivated by active studies on the PG in cuprates in terms of 
various charge fluctuations, and we have provided a microscopic basis of possible charge fluctuations in doped Mott insulators. 
We first compare our results with literature and then discuss implications for 
cuprate superconductors. 

Taking into account a number of papers about charge stripes 
in cuprates,\cite{kivelson03,vojta09} it may be surprising that 
we do not detect  stripe tendencies in our formalism, 
which  exclusively favors charge instabilities. If the $t$-$J$ model would exhibit  
a tendency to charge stripes order, it might be a consequence of 
a coupling with incommensurate magnetic modes, 
which could appear in the next-to-leading order [$O(1/N)$] in the present scheme. 
A charge-stripe solution was actually obtained in the $t$-$J$ model 
with $t'=0$ in the DMRG study,\cite{white98,white00,corboz11} which 
however contradicts other studies.\cite{hellberg99,hu12} 
In the presence of $t'$, on the other hand, 
most of numerical studies in the $t$-$J$ model 
reported that the charge stripe solution becomes unstable.\cite{white99,tohyama99} 
Our results, therefore, agree to major literature.  

Interestingly, the BOP$_{x (y)}$ shares the same feature as stripe order 
from a symmetry point of view. 
When $\vq$ shifts away from $(\pi,\pi)$, 
$\vq$ is located only along the $q_{y(x)}$ direction and thus 
the BOP$_{x(y)}$ breaks both orientational and translational symmetry of the lattice. 
In fact such an incommensurate BOP$_{x(y)}$ instability is found to occur 
up to $\delta \sim 0.10$;  see Figs.~\ref{fig2},~\ref{fig4},~and \ref{fig6}. 
While the BOP has not been discussed much so far, 
the BOP was also obtained in other studies in 
the $t$-$J$ model.\cite{cappelluti99,foussats04,morse91} 

As discussed in Sec.~III.D, 
PS at $T=0$ strongly depends on a choice of $V$, $t'$, and $J$.  
However it is a robust feature that PS occurs in an intermediate temperature 
region.  This property for a finite $T$ has not been discussed so far 
except for Ref.~\onlinecite{koch04} in the Hubbard model, probably 
because various numerical simulations are 
usually coded only at zero temperature. 
Interestingly, the reentrant critical line of PS (see the inset of \fig{fig2}) 
was interpreted as a source to generate a strong forward scattering channel of 
the electron-phonon vertex which emerges only at finite $T$.\cite{koch04} 

While the commensurate $d$PI ($\vq={\bf 0}$) was already found  in the 
$t$-$J$\cite{yamase00} and Hubbard\cite{metzner00} models, 
an incommensurate $d$PI ($\vq \ne {\bf 0}$) started to be discussed 
quite recently.\cite{metlitski10,metlitski10a,holder12,husemann12} 
We have obtained that the static $d$-wave electronic polarizability [\eq{d-bubble}] 
does not depend on $\vq$ for any momentum along the Brillouin zone diagonal, 
which holds for any temperature and any electron density as long as $t'$ is zero. 
In our model, this feature remains even for a finite $t'$ near half-filling 
since $t'$ is effectively reduced by a factor of $\delta$. 
Our result shown in the inset of \fig{fig6} (bottom panel) may be best 
compared with existing results,  since they were obtained 
in a weak coupling analysis.\cite{metlitski10,metlitski10a,holder12,husemann12} At hole density below van Hove filling ($\delta \lesssim 0.42$), 
the leading instability is an incommensurate $d$PI and its modulation vector 
is located along the $(0,0)$-$(\pi,\pi)$ direction, which 
agrees with literature.\cite{metlitski10,metlitski10a,holder12,husemann12} 
At hole density above van Hove filling we have obtained the $d$PI with $\vq={\bf 0}$,  
while Ref.~\onlinecite{holder12} showed that the static electronic polarizability 
of the $d$PI has a peak 
along the $(0,0)$-$(\pi,0)$ direction at least at $T=0$. 
In our case, an incommensurate peak along the 
$(0,0)$-$(\pi,0)$ direction indeed develops as seen in Figs.~\ref{fig3}~(b), 
\ref{fig5}~(b) for a small $t'$, but it develops 
below extremely low temperature for a large $t'$. 
This effect is not visible in \fig{fig6} (bottom). 

In the $1/N$ expansion, the $d$CDW is the leading instability in most of cases, 
in agreement with previous studies.\cite{morse91,cappelluti99,foussats04} 
We have found that close to the $d$CDW, the $d$PI also exists in a wide doping region.  
Therefore fluctuations associated with 
both $d$CDW and $d$PI are expected to be important for temperatures 
above the onset of the $d$CDW. 
The mutual interaction between $d$CDW and $d$PI  
seems rather weak at least in leading order 
because both, the full calculation [\eq{dyson}] and effective 
susceptibilities [Eqs.~(\ref{chi-c}), (\ref{chi-flux}), (\ref{chi-dPI}), and (\ref{chi-BOP})], 
give nearly the same results. 
Given that the presence of $t'$ is usually assumed for cuprates 
and our critical lines of $d$CDW and $d$PI have the same 
doping dependence of the PG temperature, furthermore with a comparable temperature 
if $t \approx 500$ meV, 
it is interesting to study each fluctuation effect on the 
electronic property. In fact,  
existing work already showed interesting results, but 
with some open questions. 
The phenomenological study assuming the $d$CDW showed that essential 
features of the PG are well captured.\cite{chakravarty01} 
However fluctuation effects were not considered in Ref.~\onlinecite{chakravarty01}. 
In a perturbative calculation of the electron self-energy due to 
a coupling to $d$CDW fluctuations in the $t$-$J$ model, 
a pseudogap is obtained in the electronic spectral function, 
which shares many important features with experimental 
data.\cite{greco09,bejas11,greco11}  
Moreover the same analysis was also applied to the 
explanation of anisotropic scattering rate of quasi-particles\cite{buzon10} 
observed in angle-dependent magnetoresistance experiments.\cite{abdel06}  
However, no calculation was performed beyond the perturbative analysis. 
On the other hand, in a perturbative calculation for $d$PI fluctuations centered 
around $\vq=(0,0)$,  
a splitting of the spectral function near the Fermi energy was obtained, reminiscent 
of a pseudogap.\cite{yamase12} 
Going beyond the perturbation theory and summing up all diagrams 
in the Gaussian fluctuation regime, however, instead of a splitting,  
the spectral function exhibits a broad single peak 
centered at the Fermi energy with a strong $\vk$ dependence of 
$d$-wave symmetry.\cite{yamase12} The spectrum in the Ginzburg 
region\cite{ginzburg60} is an open question. 
Furthermore, the  role of incommensurate $d$PI fluctuations on a pseudogap 
phenomenon remains elusive. 

The $d$PI couples directly with $xy$ anisotropy such as due to a lattice structure 
and an external strain. While the $d$PI changes to a crossover phenomenon 
in such a case, the anisotropy can be strongly enhanced by the underlying $d$PI 
fluctuations as already discussed.\cite{yamase00,okamoto10} 
The same idea is also discussed 
for iron-pnictide superconductors near the structural phase transition 
from the tetragonal to orthorhombic phase.\cite{fisher11}  
Given that lattice anisotropy 
frequently exists in cuprates, 
the relevance of the $d$PI channel in the $t$-$J$ model suggests 
important implications for understanding cuprate superconductors,  
not only for Y-based\cite{hinkov07,hinkov08,daou10,yamase06,yamase09,hackl09} 
but also for La-based\cite{yamase00,yamase01,yamase02,yamase07} compounds. 

For a large $t'$, the commensurate $d$PI appears in a heavily overdoped region 
around van Hove filling ($\delta \approx 0.3-0.45$ in \fig{fig6}). 
While our critical line exhibits reentrant behavior at low $T$,  
the canonical model for the $d$PI\cite{khavkine04,yamase05} 
suggests that the reentrant behavior is preempted 
by a first order transition as a function of the chemical potential,  
or equivalently a phase separation as a function of doping, as far as 
the $d$PI occurs at $\vq={\bf 0}$.  
It is known that Sr$_3$Ru$_2$O$_7$ exhibits the $d$PI in a magnetic 
field.\cite{grigera04,borzi07,rost09} 
In addition, a highly overdoped region in cuprates, 
where no superconducting and antiferromagnetic instabilities are expected, 
may also provide an opportunity to explore the $d$PI.

\begin{acknowledgments}
The authors thank C. Gazza, T. Holder, C. Husemann, 
W. Metzner, S. Sorella, and R. Zeyher for valuable discussions. 
M. B. thanks Consejo Nacional de Investigaciones 
Cient\'{i}ficas y T\'ecnicas (CONICET) for financial support. 
A. G. thanks the National Institute for Materials Science (NIMS), 
where this work was initiated, and the Max-Planck-Institute for hospitality.
H. Y. was supported by a Grant-in-Aid for Scientific Research from Monkasho 
and the Alexander von Humboldt Foundation. 
\end{acknowledgments}


\bibliography{main.bib}

\begin{thebibliography}{90}
\bibitem[{mis({\natexlab{a}})}]{misc-fandX}
\bibinfo{note}{Since $f$ is proportional to $X^{0p}$, $f$ and $f^{\dagger}$ do
  not fulfill the fermionic commutation rules. But after the expansion with
  respect to $\delta R$ in the Lagrangian, $f$ behaves as a fermion and its
  non-interacting propagator is given by a usual one, \eq{G0}. The price paid
  is the generation of various interaction terms between $f$ and the bosonic
  field $\delta X^{a}$. For details see Refs.~\onlinecite{foussats04} and
  \onlinecite{foussats02}.}
\bibitem[{mis({\natexlab{b}})}]{misc-foussats04}
\bibinfo{note}{The value of $V_c$ presented in Fig.~7 in
  Ref.~\onlinecite{foussats04} should be read as that of $V_c/2$.}
\end{thebibliography}

\end{document}